\documentclass[10pt]{IEEEtran}
\usepackage{times,amssymb,amsmath,amsfonts,float,nicefrac,color,bbm}
\usepackage{euscript,graphics}
\usepackage[dvips]{epsfig}
\newtheorem{theorem}{Theorem}[section]

\newtheorem{proposition}[theorem]{Proposition}

\newtheorem{cnstr}{Construction}

\newtheorem{xmpl}{Example}

\newcommand\qed{\rule{2mm}{2.5mm}\medskip}

\newcommand{\remove}[1]{}

\newcommand\nd{\noindent}

\newcommand{\ceilenv}[1]{\left\lceil #1 \right\rceil}

\newcommand\nc\newcommand
\nc\bfa{{\boldsymbol a}}\nc\bfA{{\bf A}}\nc\cA{{\mathcal A}}
\nc\bfb{{\boldsymbol b}}\nc\bfB{{\bf B}}\nc\cB{{\mathcal B}}
\nc\bfc{{\boldsymbol c}}\nc\bfC{{\bf C}}\nc\cC{{\mathcal C}}
\nc\bfd{{\boldsymbol d}}\nc\bfD{{\bf D}}\nc\cD{{\mathcal D}}
\nc\bfe{{\boldsymbol e}}\nc\bfE{{\bf E}}\nc\cE{{\mathcal E}}
\nc\bff{{\boldsymbol f}}\nc\bfF{{\bf F}}\nc\cF{{\mathcal F}}
\nc\bfg{{\boldsymbol g}}\nc\bfG{{\bf G}}\nc\cG{{\mathcal G}}
\nc\bfh{{\boldsymbol h}}\nc\bfH{{\bf H}}\nc\cH{{\mathcal H}}
\nc\bfi{{\boldsymbol i}}\nc\bfI{{\bf I}}\nc\cI{{\mathcal I}}
\nc\bfj{{\boldsymbol j}}\nc\bfJ{{\bf J}}\nc\cJ{{\mathcal J}}
\nc\bfk{{\boldsymbol k}}\nc\bfK{{\bf K}}\nc\cK{{\mathcal K}}
\nc\bfl{{\boldsymbol l}}\nc\bfL{{\bf L}}\nc\cL{{\mathcal L}}
\nc\bfm{{\boldsymbol m}}\nc\bfM{{\bf M}}\nc\cM{{\mathcal M}}
\nc\bfn{{\boldsymbol n}}\nc\bfN{{\bf N}}\nc\cN{{\mathcal N}}
\nc\bfo{{\boldsymbol o}}\nc\bfO{{\bf O}}\nc\cO{{\mathcal O}}
\nc\bfp{{\boldsymbol p}}\nc\bfP{{\bf P}}\nc\cP{{\mathcal P}}
\nc\bfq{{\boldsymbol q}}\nc\bfQ{{\bf Q}}\nc\cQ{{\mathcal Q}}
\nc\bfr{{\boldsymbol r}}\nc\bfR{{\bf R}}\nc\cR{{\mathcal R}}
\nc\bfs{{\boldsymbol s}}\nc\bfS{{\bf S}}\nc\cS{{\mathcal S}}
\nc\bft{{\boldsymbol t}}\nc\bfT{{\bf T}}\nc\cT{{\mathcal T}}
\nc\bfu{{\boldsymbol u}}\nc\bfU{{\bf U}}\nc\cU{{\mathcal U}}
\nc\bfv{{\boldsymbol v}}\nc\bfV{{\bf V}}\nc\cV{{\mathcal V}}
\nc\bfw{{\boldsymbol w}}\nc\bfW{{\bf W}}\nc\cW{{\mathcal W}}
\nc\bfx{{\boldsymbol x}}\nc\bfX{{\bf X}}\nc\cX{{\mathcal X}}
\nc\bfy{{\boldsymbol y}}\nc\bfY{{\bf Y}}\nc\cY{{\mathcal Y}}
\nc\bfz{{\boldsymbol z}}\nc\bfZ{{\bf Z}}\nc\cZ{{\mathcal Z}}
\nc\od{{\bar d}}\nc\ow{{\bar w}}\nc\odelta{{\bar\delta}}
\nc\ox{{\bar x}}\nc\oy{{\bar y}}\nc\ou{{\bar u}}
\nc\oh{{\bar h}}

\newcommand\ff{{\mathbb F}}
\newcommand\kk{{\mathbbm k}}
\newcommand\pp{{\mathbb P}}

\nc\ellone{{\ell_1}}
\nc\elltwo{{\ell_2}}
\nc\ellinf{{{\ell_\infty}}}
\nc\ip[2]{\langle #1,#2\rangle}

\newcommand{\beeq}{\begin{eqnarray*}}
\newcommand{\eneq}{\end{eqnarray*}}

\begin{document}

\sloppy

\title{Locally recoverable codes on algebraic curves} 

\author{\IEEEauthorblockN{Alexander Barg$^{a}$}\quad
\and \IEEEauthorblockN{Itzhak Tamo$^{b}$}\quad
\and \IEEEauthorblockN{Serge Vl{\u a}du{\c t}$\,^{c}$}}
\maketitle
{\renewcommand{\thefootnote}{}\footnotetext{

\vspace{-.2in}
 
\noindent\rule{1.5in}{.4pt}

\nd $^{a}$ Dept. of ECE and ISR, University of Maryland, College Park, MD 20742 and IITP, Russian Academy of Sciences, Moscow, Russia. Email abarg@umd.edu. Research supported by NSF grants CCF1422955, CCF1217894, and CCF1217245.

\nd $^{b}$ Dept. of EE-Systems, Tel Aviv University, Tel Aviv, Israel. Research done while at the Institute for Systems Research, University of Maryland, College Park, MD 20742. Email zactamo@gmail.com. Research supported by NSF grant CCF1217894.

\nd $^{c}$ Institut de Math{\'e}matiques de Marseille, Aix-Marseille Universit{\'e}, IML, 
Luminy case 907, 13288 Marseille, France, and IITP, Russian Academy of Sciences, Moscow, Russia. Email
serge.vladuts@univ-amu.fr.

}}
\renewcommand{\thefootnote}{\arabic{footnote}}
\setcounter{footnote}{0}
\thispagestyle{empty}

\vspace*{-.1in}
\begin{abstract}
A code  over a finite alphabet is called locally recoverable {(LRC code)} if every symbol in the encoding is a function of a small number (at most $r$) other symbols. A family of linear LRC codes that generalize the classic construction of Reed-Solomon codes was 
constructed in a recent paper by I. Tamo and A. Barg ({{\em IEEE Trans. Inform. Theory}, vol. 60, no.~8, 2014, pp. 4661-4676}). 
 In this paper we extend this construction to codes on algebraic curves. We give a general construction of  
LRC codes on curves and compute some examples, including asymptotically good families of codes derived from the Garcia-Stichtenoth towers. 
The local recovery procedure is performed by polynomial interpolation over $r$ coordinates of the codevector.
We also obtain a family of Hermitian codes with two disjoint recovering sets for every symbol of the codeword.
\end{abstract}
\vspace*{-.1in}
\section{Introduction: LRC Codes}
The notion of LRC codes is motivated by applications of coding to increasing reliability and efficiency of distributed storage systems.
Following \cite{gopalan2011locality}, we say that a code $\cC\subset \ff_q^n$ has locality $r$ if every symbol of the codeword 
$x=(x_1,...,x_n)\in \cC$ 
can be recovered from a subset of $r$ other symbols of $x$ (i.e., is a function of some other $r$ symbols $x_{i_1},x_{i_2},\dots,x_{i_r}$).
In other words, this means that for every $i\in [n]$ there exists a subset of coordinates 
$I_i\subset [n]\backslash i, |I_i|\le r$ such that the value of $x_i$ is be found from the restriction of $\cC$ to the coordinates in $I_i.$ 
 The subset $I_i$ is called a {\em recovering set} for the $i$th coordinate of the codeword. 

The formal definition is as follows. Given $a\in \ff_q$ consider the sets of codewords
   \begin{equation*}
   \cC(i,a)=\{x\in \cC: x_i=a\},\quad i\in[n].
   \end{equation*}
    The code $\cC$ is said to have  {\em all-symbol locality} $r$ if for every $i\in [n]$ there exists a subset $I_i\subset [n]\backslash i, |I_i|\le r$
    such that the restrictions of the sets $\cC(i,a)$ to
the coordinates in $I_i$ for different $a$ are disjoint:
 \begin{equation}\label{eq:def}
 \cC_{I_i}(i,a)\cap \cC_{I_i}(i,a')=\emptyset,\quad  a\ne a'.
 \end{equation}
We use the notation $(n,k,r)$ to refer to the parameters of the code, where $k=\log_q|C|.$


This definition can be extended to codes with {\em multiple recovering sets}; see, e.g., \cite{raw14,tam14b}. For instance, suppose that 
for every symbol $i\in[n]$ condition \eqref{eq:def} holds true for the subset $I_i, |I_i|=r_1$ as well as for some other subset $J_i\subset[n],
|J_i|=r_2.$ In this case we say that every symbol has two recovering sets, and use the notation $(n,k,\{r_1,r_2\})$ for
the code parameters. In the literature it is also often assumed that $I_i\cap J_i=\emptyset, i\in[n],$ and we include this in
the definition of LRC codes with two recovering sets (LRC(2) codes) used in this paper. 

\remove{The code $\cC_{I_i\cup \{i\}}$ is called a {\em local code} of the code $\cC$. In the constructions
of LRC codes presented in the literature the set of coordinates of the $(n,k,r)$ LRC code is usually partitioned into 
$(r+1,r)$ local MDS codes that define the recovering sets of the symbols.}

Let $\cC$ be an $(n,k,r)$ LRC code of cardinality $q^k.$ \remove{and let $R=k/n$ be the rate of $\cC.$ It is known that
\vspace*{-.1in}
  \begin{equation}
							\frac{k}{n}\leq  \frac{r}{r+1}.
							\label{eq:rate}
				\end{equation}}
				The minimum distance of $\cC$ is known to
				satisfy \cite{gopalan2011locality,pap12}
   			\begin{equation}
				d\leq n-k-\ceilenv{ \frac{k}{r}}+2.				 
				\label{eq:dist}
     		\end{equation} 
\vspace*{-.05in}

The bound \eqref{eq:dist} is an extension of the classical Singleton bound of coding theory, which is attained by the 
well-known family of Reed-Solomon (RS) codes. 
RS-like codes with the LRC property whose parameters meet the bound \eqref{eq:dist} were recently constructed in \cite{tam14a}.
Unlike some other known constructions, e.g., \cite{Natalia,My-paper}, the codes \cite{tam14a} are constructed over finite fields of cardinality comparable to the code length $n.$ 
The cyclic case of the construction in \cite{tam14a} is studied in the recent paper \cite{tam15a}.

Classical RS codes can be viewed as a special case of the general construction of geometric Goppa codes; in particular, good codes
are obtained from families of curves with many rational points \cite{TVN07}. Motivated by this approach,
in this paper we take a similar view of the construction of the evaluation codes of \cite{tam14a}.
Interpreting these codes as codes on algebraic curves, we present a general construction of 
algebraic geometric LRC codes and compute the parameters of codes for some particular choices
of curves. Similarly to \cite{tam14a}, local recovery can be performed by interpolating a univariate polynomial over $r$ coordinates
of the codeword. The RS-like codes in \cite{tam14a} can be extended to multiple recovering sets, and here we point out one
such extension in the case of Hermitian codes. 

\remove{In classical coding theory, there is also another line of thought associated with RS codes. Confining oneself to the cyclic case,
we can phrase the study of code parameters in terms of the zeros of the code. Importantly, this point of view leads to a number
of nontrivial results for {\em subfield subcodes} of RS codes such as BCH codes and related code families. A similar study can
be performed for LRC RS codes, with the main outcome being a characterization of both the distance and locality in terms of the zeros
of the code. This point of view is further developed in \cite{tam15a}.}

\vspace*{-.1in}

\section{The construction of \cite{tam14a}}
Let us briefly recall the construction of \cite{tam14a}.
Our aim is to construct an LRC code over $\ff_q$ with the parameters $(n,k,r)$, 
where $n\le q.$ We additionally assume that $(r+1)|n$ and $r|k$, although both the constraints can be lifted
by adjustments to the construction presented below \cite{tam14a}. Let $g(x)\in \ff_q[x]$ be
a polynomial of degree $r+1$ such that there exists a partition $\cA=\{A_1,\dots,A_{\frac n{r+1}}\}$
of a set of points $A=\{P_1,\dots,P_n\}\subset \ff_q$ into subsets of size $r+1$ such that $g$ is constant on each set $A_i\in \cA.$

Consider a $k$-dimensional linear subspace $V\subset \ff_q[x]$ generated by the set of polynomials
   \begin{equation}\label{eq:basis}
     (g(x)^j x^i,\; i=0,\dots, r-1; j=0,\dots,\frac kr-1).
   \end{equation}
Given $a=(a_{ij},i=0,\dots,r-1;j=0,\dots,\frac kr-1)\in \ff_q^k$
    let 
   \begin{multline}
   f_a(x)=\sum_{i=0}^{r-1}f_i(x)x^i, \text{ where } f_i(x)=\sum_{j=0}^{\frac kr-1} a_{ij}g(x)^j,
   \\ i=0,\dots,r-1. \label{eq:fa}
   \end{multline}
Now define the code $\cC$ as the image of the linear evaluation map
  \begin{equation}\label{eq:cc}
  \begin{aligned}
    e:V&\to\ff_q^n\\
    f_a&\mapsto (f_a(P_i), i=1,\dots,n).  
  \end{aligned}
  \end{equation}
As shown in \cite{tam14a}, $\cC$ is an $(n,k,r)$ LRC code whose minimum distance $d$ meets the bound \eqref{eq:dist}
with equality.  

To construct examples of codes using this approach we need to find polynomials and partitions of points of 
the field that satisfy the above assumptions. As shown in \cite{tam14a}, one can take $g(x)=\prod_{\beta\in H}(x-\beta),$ where $H$ is a subgroup of the additive or the multiplicative group of $\ff_q$ (see also the 
example in the next section).
In this case $r=|H|-1,$ and the corresponding set of points $A$ can be taken to be any collection of the cosets
of the subgroup $H$ in the full group of points. In this way we can construct codes of length
$n=m(r+1),$ where $m\ge 1$ is an integer that does not exceed $(q-1)/|H|=(q-1)/(r+1)$ or $q/|H|{=q/(r+1)}$ depending on the choice
of the group. 
  
\vspace*{-.1in}\section{Algebraic geometric LRC codes}
As above, let us fix a finite field $\kk ={\ff}_q, q=p^a$ of characteristic $p$. To motivate our construction,
consider the following example.

{\em Example 1:} Let $H$ be a cyclic subgroup of $\ff_{13}^\ast$ generated by $3$ and let $g(x)=x^3.$ Let $r=2,n=9,k=4,$ and choose
$A=\{1,2,3,4,5,6,9,10,12\}.$ We obtain
  $\cA=\{A_1,A_2.A_3\},$ where
  \begin{equation}\label{eq:PP}
  \begin{aligned}
  A_1=\{1,3,9\},\;&&A_2=\{2,6,5\},\;&&A_3=\{4,12,10\}\\
      g(A_1)=1 & &g(A_2)=8 &&g(A_3)=12
  \end{aligned}
  \end{equation}
The set of polynomials \eqref{eq:basis} has the form $(1,x,x^3,x^4).$
In this case Construction \eqref{eq:cc} yields a $(9,4,2)$ LRC code with distance $d=5$ \cite{tam14a}.

This construction can be given the following geometric interpretation: the polynomial $g$ defines a mapping $g:{\mathbb P}^1 \to {\mathbb P}^1$ of degree $r+1=3$ such that the preimage of every point in $g(A)$ consists of
``rational'' points (i.e., $\ff_q$-points). This suggests a generalization of the construction to algebraic curves which we proceed
to describe (note Example 2 on the following page that may make it easier to understand the general case).

Let $X$ and $Y$ be smooth projective absolutely irreducible curves over $\kk$. Let $g:X\to Y$ be a rational
separable map of curves of degree $r+1.$ As usual, denote by $\kk(X)$ ($\kk(Y)$) the field of rational functions on $X$ (resp., $Y$).
Let $g^\ast:\kk(Y)\to \kk(X)$ be the function that acts on $\kk(Y)$ by
$g^\ast(f)(P)=f(g(P)),$ where $f\in \kk(Y), P\in X.$  The map $g^\ast$ defines a field embedding $\kk(Y)\hookrightarrow \kk(X),$ and we identify $\kk(Y)$ with its image $g^*( \kk(Y))\subset \kk(X).$  

Since $g$ is separable, the primitive element
theorem implies that there exists a function $x\in \kk(X)$ such that $\kk(X)=\kk(Y)(x)$, and that satisfies
the equation 
  $$
  x^{r+1}+b_r x^r+\dots+b_0=0,
  $$
  where $b_i\in \kk(Y).$
The function $x$ can be considered as a map $x: X\rightarrow \mathbb{P}^1_\kk,$ and we denote its degree $\deg (x)$ by $h.$ 

{\em Example 1:} (continued) For instance, in the above example, we have $X={\mathbb P}^1, Y={\mathbb P}^1,$ and the mapping $g$ is given by $y=x^3.$ We obtain $\kk(Y)=\kk(x^3)=\kk(y),$ $\kk(X)=\kk(y)(x),$ where $x$ satisfies the equation 
$x^3-y=0.$ Note that in this case $b_r=b_{r-1}=...=b_1=0,b_0=y.$

The codes that we construct belong to the class of evaluation codes. Let 
$S=\{P_1,\dots,P_s\}\subset Y(\kk)$ be a subset of $\ff_q$-rational points of $Y$ and let $Q_\infty$ be a 
positive divisor of degree $\ell\ge 1$ whose support is disjoint from $S.$ For instance, one can assume that $Q_\infty=\pi^{-1}(\infty)$ for
a projection $\pi:Y\to \mathbb{P}^1_\kk.$ To construct our 
codes we introduce the following set of fundamental assumptions with respect to $S$ and $g$:
    \begin{gather}
  \hspace*{-.1in}A:= g^{-1}(S)=\{P_{ij}, i=0,\dots, r, j=1,\dots,s\}\subseteq X(\kk);\label{eq:partition} \\
    g(P_{ij})=P_j \text{ for all } i,j; \notag \\
    b_i\in  L(n_iQ_{\infty}), \quad i=0,1,\dots,r, \notag
    \end{gather}
for some natural numbers $n_i.$ 

Now let $D=tQ_{\infty}$ be a positive divisor and let $\{f_1,\ldots, f_m\}$ be a basis of the linear space $L(D).$ 
The functions $f_i, i=1,\dots,m$ are contained in $\kk(Y)$ and therefore are constant on the fibers of the map $g$.
The Riemann-Roch theorem implies that $m\ge  t\ell-g_Y+1,$ where $g_Y$ is the genus of $Y.$
Below we assume that $m=  t\ell-g_Y+1.$
  Consider the $\kk$-subspace $V$ of $\kk(X)$ of dimension $rm$ generated by the functions
  \begin{equation}\label{eq:fs}
 \{f_jx^i, i=0,\ldots,r-1,j=1,\ldots,m\}
  \end{equation}
  (note an analogy with \eqref{eq:basis}). Since $Q_{\infty}$ is disjoint from $S$, the evaluation  map 
  \begin{equation}\label{eq:eval}
  \begin{aligned}
   e:=ev_A: &V\longrightarrow \kk^{(r+1)s}\\ 
     &F\mapsto (F(P_{ij}),i=0,\dots, r, j=1,\dots,s)
   \end{aligned}
  \end{equation}
 is well-defined. The image of this mapping is a linear subspace of $\ff_q^{(r+1)s}$ (i.e., a code), which we denote 
by $\cC(D,g).$ The code coordinates are naturally partitioned into $s$ subsets $A_j=\{P_{ij},i=0,...,r\}, j=1,\dots,s$ of size $r+1$ each; see \eqref{eq:partition}. Assume throughout that, for any fixed $j$,  $x$ takes different values at the points in the set $(P_{ij},i=0,\dots,r).$
\begin{theorem}\label{thm:LRC} The subspace $\cC(D,g)\subset \ff_q$ forms an $(n,k,r)$ linear LRC code with the parameters
   \begin{equation}
\left.\begin{array}{c}
 n=(r+1)s  \\[.05in] k=rm\ge r(t\ell-g_Y+1) \\ [.05in]
 d\ge n-t\ell(r+1)-(r-1)h\end{array}\right\} \label{eq:d}
 \end{equation}
provided that the right-hand side of the inequality for $d$ a positive integer.  Local recovery of an erased symbol $c_{ij}=F(P_{ij})$
can be performed by polynomial interpolation through the points of the recovering set $A_j$.
\end{theorem}
\begin{IEEEproof}
The first relation in \eqref{eq:d} follows by construction. The inequality for the distance is also immediate:
the function $f_j x^i,f_j\in L(D)$, evaluated on $A,$  can have at most $t\ell(r+1)+(r-1)\deg (x)$ zeros. 
Since we assume that $d\ge1,$ the mapping $ev_T$ is injective, which implies the claim about the dimension of the code.
Finally, the functions $f_i$ are constant on the fibers $(P_{ij},i=0,\dots,r-1);$ therefore on each subset $A_j$ the
codeword is obtained as an evaluation of the polynomial of the variable $x$ of degree $\rho\le r-1.$ This representation accounts
for the fact that coordinate $c_P, P\in A_j$ of the codeword can be found by interpolating a polynomial of degree at most $r_1$ through the remaining points of $A_j.$
\end{IEEEproof}

 \vspace*{-.1in}\section{Some code families} 
 Let us give some examples of code families arising from our construction.
 
 \subsection{LRC codes from Hermitian curves}\label{sect:Herm}
 Let $q=q_0^2,$ where $q_0$ is a prime power, let $\kk=\ff_q,$ and
let $X:=H$ be the Hermitian curve, i.e., a plane smooth curve of genus $g_0=q_0(q_0-1)/2$ with the affine equation
     $$
     X: x^{q_0}+x=y^{q_0+1}.
     $$
     The curve $X$ has $q_0^3+1=q\sqrt q+1$ rational points of which one is the infinite point and the remaining
$q_0^3$ are located in the affine plane. There are two slightly different ways of constructing Hermitian LRC codes.
 
 \subsubsection{\underline{Projection on $y$}}
Here we construct $q$-ary $(n,k,r=q_0-1)$ LRC codes.
Take $Y={\mathbb P}^1(\kk)$ and take $g$ to be  the natural projection defined by $g(x,y):=y,$
then the degree of $g$ is $q_0=r+1$ and the degree of $x$ is $h=q_0+1.$ We can write
$X(\ff_q)=g^{-1}(\ff_q)\bigcup Q_{\infty}'$ where $Q_{\infty}'\in X$ is the  unique point over $ {\infty}\in Y.$

Turning to the code construction, take $ S=\kk\subset \mathbb{P}^1, Q_{\infty}={\infty},\ell=1,$ 
and $D=tQ'_\infty$ for some $t\ge 1.$ We have
   $$
   L(D)=\Big\{\sum_{i=0}^t a_i y^i\Big\} \subset \kk[y].
   $$
 Following the general construction of the previous section, we obtain the following result.
 \begin{proposition} \label{prop:Herm}The construction of Theorem \ref{thm:LRC} gives a family of $q$-ary Hermitian LRC codes
 with the parameters
   \begin{gather*}
   n=q_0^3, k=(t+1)(q_0-1), r=q_0-1\\
  d\ge n-tq_0-(q_0-2)(q_0+1).
  \end{gather*}
  \end{proposition}
{\em Example 2:} Let $q_0=3,q=9,\kk=\ff_9$ and consider the Hermitian curve $X$ of genus 3 given by the equation
    $
    x^3+x=y^4.
    $
The curve $X$ has 27 points in the finite plane, shown in Fig.1 below (here $\alpha^2=\alpha+1$ in $\ff_9$), and one point at infinity.
 \remove{ \begin{align}
   \;&(0,0),(\alpha^2,0),(\alpha^6,0); \notag\\
  &(\alpha,\beta),(\alpha^3,\beta),(\alpha^4,\beta), \;\beta=1,\alpha^2,\alpha^4,\alpha^6; \label{eq:points}\\
  &(1,\beta),(\alpha^5,\beta),(\alpha^7,\beta),\;\beta=\alpha,\alpha^3,\alpha^5,\alpha^7,\notag
  \end{align}
where $\alpha^2=\alpha+1$ in $\ff_9,$ and one point at infinity. }
    \nc{{\bl}}{\bullet}
\vspace*{-.05in}\hrule   \begin{gather*}
  \begin{array}{c@{\hspace*{.05in}}c@{\hspace*{.05in}}*{10}{c@{\hspace*{.05in}}}}
   &\alpha^7&    &    &\bullet  &          &\bullet   &        &\bullet  &        &\bullet&\\  
   &\alpha^6&\bl &    &         &          &          &        &         &        &       &\\
   &\alpha^5&    &    &\bl      &          &\bl       &        &\bl      &        &\bl    &\\
   &\alpha^4&    &\bl &         &\bl       &          &\bl     &         &\bl     &       &\\
 x &\alpha^3&    &\bl &         &\bl       &          &\bl     &         &\bl     &       &\\
   &\alpha^2&\bl &    &         &          &          &        &         &        &       &\\
   &\alpha&      &\bl &         &\bl       &          &\bl     &         &\bl     &       &\\
   &1       &    &    &\bl      &          &\bl       &        &\bl      &        &\bl    &\\
   &0       &\bl &    &         &          &          &        &         &        &       &\\
     &&0 &1 &\alpha &\alpha^2 &\alpha^3 &\alpha^4 &\alpha^5 &\alpha^6 &\alpha^7&\\
     &&  &  &       &         &         &y
  \end{array}\\
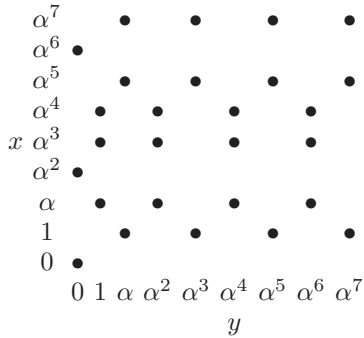
 \text{\footnotesize Fig.1: 27 points of the Hermitian curve over $\ff_{9}.$ }
 \end{gather*}
\hrule
\vspace*{.1in} \nd
The columns of the array in Fig. 1 correspond to the fibers of the mapping $g(\cdot,y)=y,$ 
and for every $a\in Y(\ff_9)\backslash Q_\infty$ there are 3 points $(\cdot,a)\in X$ lying above it.
These triples form the recovering sets $A_1,\dots,A_9,$ similarly to \eqref{eq:PP}.
  The map $x:X\to{\mathbb P}^1$ has degree $h=4.$ Choosing $D$ in the form $D=tQ'_\infty$ and taking $S=\ff_9$ 
(all the affine points of $Y$), we obtain an LRC code $\cC(D,g)$ with the parameters 
   \begin{gather}
    n=27, k=2(t+1), r=2\\
     d\ge 27-3t-4=23-3t.
     \end{gather}
 
For instance, take $t=2.$ The basis of functions \eqref{eq:fs} in this case takes the following form:
  \begin{equation*}
  \{1,y,y^2,x,xy,xy^2\}.
  \end{equation*}
To give an example of local decoding, let us compute the codeword for the message vector $(1,\alpha,\alpha^2,\alpha^3,\alpha^4,\alpha^5).$ 
The polynomial
$$   
  F (x,y)=1+\alpha y+\alpha^2y^2+\alpha^3x+\alpha^4xy+\alpha^5xy^2
  $$
evaluates to the codeword
\begin{gather*}
  \begin{array}{c@{\hspace*{.05in}}c@{\hspace*{.1in}}*{10}{c@{\hspace*{.05in}}}}
   &\alpha^7&        &    &\alpha   &          &\alpha^7  &        &\alpha^5 &         &0     &\\  
   &\alpha^6&\alpha^2&    &         &          &          &        &         &         &      &\\
   &\alpha^5&        &    &\alpha^6 &          &\alpha^4  &        &\alpha^2 &         &0     &\\
   &\alpha^4&        &\alpha^7&     &\alpha^3  &          &\alpha^5&         &\alpha^5 &      &\\
x  &\alpha^3&        &\alpha^3&     &\alpha^7  &          &\alpha  &         &\alpha   &      &\\
   &\alpha^2&\alpha^3&    &         &          &          &        &         &         &      &\\
   &\alpha  &        &0   &         &0         &          &0       &         &0        &      &\\
   &1       &        &    &1        &          &\alpha^6  &        &\alpha^4 &         &0     &\\
   &0       &1       &    &         &          &          &        &         &         &      &\\[.05in]
          &&0       &1   &\alpha   &\alpha^2  &\alpha^3 &\alpha^4 &\alpha^5 &\alpha^6 &\alpha^7&\\
            &&  &  &       &         &         &y\vspace*{-.05in}
  \end{array}
  \end{gather*}
(e.g., $F(0,0)=1$, etc.).
Suppose that the value at $P=(\alpha,1)$ is erased. The recovering set for the coordinate $P$ is $\{(\alpha^4,1),(\alpha^3,1)\},$
so we compute a linear polynomial $f(x)$ such that $f(\alpha^4)=\alpha^7$ and $f(\alpha^3)=\alpha^3,$ i.e., 
$f(x)=\alpha x-\alpha^2.$ Now the coordinate at $(\alpha,1)$ can be found as $f(\alpha)=0.$\hfill\qed
 
\vspace*{-.05in}  Computing the gap to the Singleton bound \eqref{eq:dist}, we obtain
   \begin{align}
   d+\frac kr(r+1)&\ge q_0^3-tq_0-(q_0-2)(q_0+1)+q_0(t+1)\nonumber\\
       &=q_0^3-q_0^2+2q_0+2\nonumber\\
       &=n-q+2\sqrt q +2.\label{eq:sg}
       \end{align}
 For codes that meet the bound \eqref{eq:dist} we would have $d+k(r+1)/r=n+2,$
 so the Singleton gap of the Hermitian LRC codes is {at most} $q-2\sqrt q=q_0(q_0-2).$ Of course, these codes cannot be Singleton-optimal because their length 
 is much greater than the alphabet size, but the gap in this case is still rather small.
 For instance in Example 2 we have $d+k(r+1)/r\ge 23-3t+3(t+1)=26$ while for codes meeting the Singleton bound we would have 
 $d+k(r+1)/r=29.$ 

\subsubsection{\underline{Projection on $x$}}\label{sect:Herm2} Again take $Y=\pp^1$ and let $g'(x,y):=x$ be the second natural projection on $\pp^1.$
There are $q_0$ points on $\pp^1$ that are fully ramified (they have only one point of $X$ above them), namely the points in the set
  \begin{equation}\label{eq:M}
  M=\{a\in \ff_q: a^{q_0}+a=0\}
  \end{equation}
(e.g., in Fig.~1 $M=\{0,\alpha^2,\alpha^6\}$). Therefore, every fiber of $g'$ over $\ff_q\backslash M$ consists of $\ff_q$-rational
points since there are in total
   $$
   |\ff_q\backslash M|\cdot(q_0+1)=q_0^3-q_0
   $$
rational points in those fibers.    Obviously $|g^{-1}(a)\cap g'^{-1}(b)|\le1$ for all $a,b\in \ff_q.$

Take $S=\ff_q\backslash M$, then $r=q_0,$ and clearly $h=\deg(y)=q_0.$
We obtain
  \begin{proposition}
  The construction of Theorem \ref{thm:LRC} gives a family of $q$-ary Hermitian LRC codes
 with the parameters
   \begin{gather*}
   n=q_0^3-q_0, k=(t+1)q_0, r=q_0\\
  d\ge n-t(q_0+1)-q_0(q_0-1).
  \end{gather*}
  \end{proposition}
  For instance, in Example 2, taking $t=2,$ we obtain a code of dimension 9 from the basis of functions
  $
    \{1,y,y^2,x,xy,xy^2,x^2,x^2y,x^2y^2\}.
    $

Performing a calculation similar to \eqref{eq:sg} we obtain the quantity one less than for the first family:
   $$
   d+\frac kr(r+1)=n-q+2\sqrt q+1.
   $$
   
{\em Remark 4.1:} Hermitian LRC codes are in a certain sense optimal for our construction. Note that 
most known curves with the optimal quotient (number of rational points)/(genus) have the property that
for any projection $g:X\to \pp^1$ the point $\infty\in \pp^1$ is totally
ramified (see e.g., the next section). In this case the quantity $h$ satisfies $h\ge n/q.$ At the same time, for
Hermitian curves, $h=n/q$ (or $(n/q)+1$). Recall also that Hermitian curves are absolutely maximal,
i.e. attain the equality in Weil's inequality, and moreover, their genus is maximal for
maximal curves.

\subsubsection{\underline{ Two recovering sets}} The existence of two projections $g$ and $g'$ with mutually transversal fibers suggests
that Hermitian LRC codes could be modified, leading to a family of LRC(2) codes
with two recovering sets of size $r_1=q_0-1$ and $r_2=q_0,$ respectively.
Indeed, let
   $$
   B=g^{-1}(\ff_q\backslash\{0\})=(g')^{-1}(\ff_q\backslash M)\subset X/\ff_q ,
   $$
$|B|=(q_0^2-1)q_0,$ where $M$ is defined in \eqref{eq:M}, and consider the following polynomial space of dimension $(q_0-1)q_0:$
   $$
   L:=\text{span\,}\{x^iy^j, i=0,1,\dots,r_1-1,j=0,1,\dots,r_2-1\}.
   $$
\begin{proposition} Consider the linear code $\cC$ obtained by evaluating the functions in $L$ at the points of $B$.
  The code $\cC$ has the parameters $(n=(q_0^2-1)q_0,k=(q_0-1)q_0,\{r_1=q_0-1,r_2=q_0\})$ and distance
    \begin{equation}\label{eq:distance}
      d\ge (q_0+1)(q_0^2-3q_0+3).
    \end{equation}
\end{proposition}
  \begin{IEEEproof} $X$ is a plane curve of degree $q_0+1$, so the Bezout theorem implies that any polynomial of degree $\le 2q_0-3$
  has no more than $(q_0+1)(2q_0-3)$ zeros on $X;$ hence \eqref{eq:distance}.\end{IEEEproof}
For instance, puncturing the code of Example 2 on the coordinates in $M,$ we obtain an LRC(2) code with the parameters
$(24,6,\{2,3\})$ and distance $d\ge 12.$

\vspace*{-.1in} 
\subsection{Codes from Garcia-Stichtenoth curves}
 Let $q=q_0^2$ be a square and let $l\ge 2$ be an integer. Define the curve $X_l$ and the functions $x_l,z_l$ inductively as follows:
    \begin{gather}
 x_0:=1;\;X_1:= \mathbb{P}^1, \kk(X_1)=\kk(x_1); \label{eq:GST1}\\
X_l: z_l^{q_0}+z_l=x_{l-1}^{q_0+1}, x_{l-1}:=\frac{z_{l-1}}{x_{l-2}} \in \kk(X_{l-1}) \text{ (if } l\ge 3),\notag
  \end{gather}
  where $\kk=\ff_q.$
In particular, $X_2=H$ is the Hermitian curve.
 The resulting family of curves is known to be asymptotically maximal \cite{garcia95}, \cite[p.177]{TVN07}, 
 and gives rise to codes with good parameters in the standard error correction problem. 
Since this family generalizes Hermitian curves, we can expect that it gives rise to two families of codes that extend
the constructions of Sect.~\ref{sect:Herm}. This is indeed the case, as shown below.

\subsubsection{}\label{sect:GS1}
To use the general construction that leads to Theorem \ref{thm:LRC} we take the map $g_l:X_l\to X_{l-1}$ to be the natural projection  of
degree $q_0=r+1.$ We note that
    \begin{equation}\label{eq:gs1}
        g_l^\ast: \kk(X_{l-1})\to\kk(X_l)=\kk(X_{l-1})(z_l).
    \end{equation}
To describe rational points of the curve $X_l$ let $\psi_l:X_l\to {\mathbb P}^1$ be the natural projection of degree $q_0^{l-1},$
i.e., the map $\psi_l=g_l\circ g_{l-1}\circ\dots\circ g_2.$ Then all the points in the preimage $P_l:=\psi_l^{-1}(\ff_q^\ast)$ are
$\ff_q$-rational, and there are $n_l=q_0^{l-1}(q_0^2-1)$ such points.
The genus of the curve $X_l$ can be bounded above as
     $$
     G_l\le q_0^{l}+q_0^{l-1}=q_0^{l-1}(q_0+1)=\frac{n_l}{q_0-1}
     $$
(the exact value of $G_l$ is known \cite{garcia95}, but this estimate suffices: in particular, it implies that the curves $X_l,\l\to \infty$
are asymptotically maximal). 
We obtain the following result.
 \begin{proposition}\label{prop:GS1}  There exists 
 a family of $q$-ary $(n,k,r=q_0-1)$ LRC codes on the curve $X_l, l\ge 2$  with the parameters
   \begin{equation}\label{eq:GS1}
   \left.
   \begin{array}{c}
   n=n_l=q_0^{l-1}(q_0^2-1)\\[.1in]
   \displaystyle k\ge r\Big(t-\frac{n_{l-1}}{q_0-1}+1\Big) \\[.1in]
  \displaystyle d\ge n_l-t q_0-\frac{2n_l(q_0-2)}{q_0^2-1}
  \end{array}\right\}
  \end{equation}
  where $t$ is any integer such that $G_{l-1}\le t\le n_{l-1}.$
  \end{proposition}
  \begin{IEEEproof}
We apply the construction of Theorem \ref{thm:LRC} to $X:=X_l, Y:=X_{l-1},$ taking the map $g:=g_l, $ $Q_\infty:=P_{\infty,l},$ $D=tQ_{\infty},$ and $\ell=1.$

The function $x$ in the general construction in this case is $x=z_l.$ To estimate the distance of the code $\cC(D,g)$ using \eqref{eq:d} we
need to find the degree $h=\deg(z_l).$ Toward this end, observe that $z_l=x_lx_{l-1},$ so 
   $$
    \deg(z_l)=\deg(x_l)+\deg(x_{l-1}).
    $$
Let $(x_l)_0^{(l)}$ be the divisor of zeros of $x_l$ on $X_l.$ Recall from \cite{garcia95}, Lemma 2.9 that  $(x_l)_0^{(l)}=q_0^{l-1}Q_l,$
where $Q_l$ is the unique common zero of $x_1,z_2,\dots,z_l.$ Therefore, $\deg(x_l)_0^{(l)}=q_0^{l-1}$ and $\deg(x_{l-1})_0^{(l-1)}=q_0^{l-2}.$
Since the map $X_l\to X_{l-1}$ is of degree $q_0,$ we obtain $\deg (x_{l-1})_0^{(l)}=q_0\deg(x_{l-1})_0^{(l-1)}=q_0^{l-1}.$
Summarizing, 
 $$
   h=2q_0^{l-1}=\frac{2n_{l}}{q_0^2-1}.
 $$
Now the parameters in \eqref{eq:GS1} are obtained from \eqref{eq:d} by direct computation. 
\end{IEEEproof}
 
\subsubsection{}\label{sect:GS2} Now consider the second natural projection of curves in the tower \eqref{eq:GST1}. Namely, let
$Y_l$ correspond to the function field $\kk(z_2,\dots,z_l)$ and consider the field embedding
   $$
      (g_l')^\ast:\kk(Y_l)\to \kk(X_l)=\kk(x_1,z_2,\dots,z_l).
      $$
Note that $g_2'$ is the projection $g':X_2\to\pp^1$ considered in Section \ref{sect:Herm2}. The curves
$\{Y_l,l=2,3,\dots\}$ form another optimal tower of curves \cite[Remark 3.11]{garcia96} given by the recursive equations
   $$ 
   Y_{l}: z_{l}^q +z_{l} =\frac{z_{l-1}^q }{z_{l-1}^{q-1} + 1}, \;l\ge 3;\; 
Y_2:=\mathbb{P}^1.
   $$
In geometric terms, the embedding $(g'_l)^\ast$ implies that 
the curve $X_l$ is the fiber product of $X_2$ and $Y_l$ over $Y_2=\mathbb{P}^1,$ viz.
$X_l= X_2\times_{Y_2}Y_l,$ which in turn implies that the projection 
$g'_l: X_{l}\rightarrow  Y_l$ shares the main properties of $g'=g'_2$. Indeed, we have:
\begin{enumerate}
  \item The genus of $Y_l$ satisfies $G_l'<q_0^{l-1}$ (the exact value is given in \cite[Remark 3.8]{garcia96}; note that the
  notation for $\kk(Y_l)$ in \cite{garcia96} is $T_{l-1}).$
  \item Let $\pi_l:Y_l\to Y_2$ be the natural projection of degree $\deg(\pi_l)=q_0^{l-2}$. All the points in $S_l:=\pi_l^{-1}(\ff_q\backslash
  M)$ are $\ff_q$-rational and
     $$
  |S_l|=q_0^{l-2}(q_0^2-q_0)=q_0^{l-1}(q_0-1)=n_l/(q_0+1).
     $$
 \item The point $\infty\in Y_2=\pp^1$ is totally ramified, i.e., $\pi_l^{-1}(\infty)=P_{\infty,l}'$
     for a rational point $P_{\infty,l}'\in Y_l.$
 \item We have $(g'_l)^{-1}(S_l)= (\psi_l)^{-1}(\ff_q), |(g'_l)^{-1}(S_l)|=n_l,$ and all the points in $(g'_l)^{-1}(S_l)$ are $\ff_q$-rational.  The degree of the projection $g_l'$ is $ \deg (g'_l)=q_0+1.$ The fibers of $g'_l$ are transversal with those of  $g_l.$
\item The degree of $x_1: X_l\longrightarrow \mathbb{P}^1$ equals 
$h:=\deg (x_1)= \deg (\pi_l)\deg ((x_1)_0^{(2)})=q_0^{l-1}.$          
  \end{enumerate}
We obtain the following statement.
\begin{proposition}\label{prop:GS2}
There exists  a family of $q$-ary $(n,k,r=q_0)$ LRC codes on the curve $X_l, l\ge 2$  with the parameters
   \begin{equation}\label{eq:GS2}
   \left.
   \begin{array}{c}
   n=n_l=q_0^{l-1}(q_0^2-1)\\[.1in]
   \displaystyle k\ge r(t-q_0^{l-1}+1) \\[.1in]
  \displaystyle d\ge n_l-t (q_0+1)-(q_0-1)q_0^{l-1}
  \end{array}\right\}
  \end{equation}
  where $t$ is any integer such that $G_{l-1}\le t\le n_{l-1}.$
\end{proposition}  
{\em Proof:} Put $r=q_0,\ell=1$ and apply the construction of Theorem \ref{thm:LRC} to 
   $$
X:=X_l, Y:=Y_l, g:=g_l', Q_\infty:=P_{\infty,l}', D_t=tQ_{\infty}.\hspace*{.3in}\qed
   $$
   
{\em Remark 4.2:} For the construction of Prop.~\ref{prop:GS2} the lower bound of Remark 4.1
takes the form $h\ge n_l/q_0^2=q_0^{l-1}-q_0^{l-3}$ which is very close the actual value $h=q_0^{l-1}.$
In the case of Prop. \ref{prop:GS1} the value $h$ is about twice as large as the lower bound. 

{\em Remark 4.3:} Due to the results of \cite{shum01}, the basis of the function space $L(D_t)$ and the set $S_l$ can be
found in time polynomial in $n_l$, and so the codes of Prop.~\ref{prop:GS2} are polynomially constructible.
 
\vspace*{-.15in}  
\subsection{Asymptotic bounds} 
Let us compute the asymptotic relation between the parameters of LRC codes on the 
Garcia-Stichtenoth curves constructed above.
\begin{proposition} \label{prop:ab}
Let $q=q_0^2,$ where $q_0$ is a prime power. There exist families of LRC codes with locality $r$
whose rate and relative distance satisfy
   \begin{align}
     R&\ge \frac r{r+1}\Big(1-\delta-\frac 3{\sqrt q+1}\Big), && r=\sqrt q-1   
    \label{eq:ab1}\\
      R&\ge \frac r{r+1}\Big(1-\delta-\frac{2\sqrt q}{q-1}\Big),&& r=\sqrt q . \label{eq:ab2}
     \end{align}
 \end{proposition}
{\em Remark 4.4:} Recall that without the locality constraint the relation between $R$ and $\delta$ for codes on asymptotically
optimal curves (for instance, on the curves $X_l,l=2,3,\dots$) takes the form 
  $
  R \ge 1-\delta-\frac{1}{\sqrt q-1},
  $ see \cite[p.251]{TVN07}.
 \begin{IEEEproof}
For instance, let us check \eqref{eq:ab1}. From \eqref{eq:GS1} we obtain
  \begin{align*}
  d+\frac{k(r+1)}{r}&\ge n_l-\frac{q_0n_{l-1}}{q_0-1}-\frac{2n_l(q_0-2)}{q_0^2-1}+q_0\\
  &\ge n_l\Big(1-\frac{1}{q_0-1}-\frac{2q_0-4}{q_0^2-1}\Big)\\
  &=n_l\Big(1-\frac{3}{q_0+1}\Big).
  \end{align*}
Letting $\delta=d/n_l, R=k/n_l, l\to\infty,$ we obtain \eqref{eq:ab1}.
\end{IEEEproof}

One of the main results about the classic AG codes is an improvement of the Gilbert-Varshamov (GV) bound starting with $q=49.$  
A GV-type bound for LRC codes was recently obtained in \cite{BFT}.
\begin{theorem} There exists a sequence of $q$-ary linear $r$-LRC codes with the parameters $(R,\delta)$ as long as 
\begin{equation}\label{eq:GV}
    R< \frac{r}{r+1} - \min\limits_{0<s\le 1} \Big\{ \frac{1}{r+1}\log_q b(s) - \delta \log_q s \Big\},
    \end{equation}
 where
 \begin{equation}\label{eq0:fg}
    b(s)  = \frac{1}{q}( ( 1 + (q-1)s)^{r+1} + (q-1) (1-s)^{r+1}).
    \end{equation}
\remove{as long as
  \begin{equation}\label{eq:GV}
  R<\frac r{r+1}-\delta\log_q\frac{q-1}{\delta}-(1-\delta)\log_q\frac{1}{1-\delta}
  \end{equation}
  for all $\delta$ such that the right-hand side is positive.}
    \end{theorem}
\remove{    
    \begin{IEEEproof}
Construct a linear code $C'\subset \ff_q^n$
with distance $d$ by successively adding columns to the parity-check matrix. A code of length $n$ and dimension
$k'$ exists if
    $$
     \sum_{i=0}^{d-2} \binom {n-1}i(q-1)^i< q^{n-k'}.
     $$
     To account for the LRC condition, add $\lceil n/(r+1)\rceil$ parity equations that provide the local recovery
     property. This results in an $(n,k,r)$ LRC code with distance $d$ and dimension
     $$
     k\ge k'-\Big\lceil\frac n{r+1}\Big\rceil.
     $$
Letting $n\to\infty,$ $R=k/n$ and $\delta=d/n,$ we obtain \eqref{eq:GV}.
\end{IEEEproof}}
The bound given in \eqref{eq:ab2} (i.e., the code family constructed in Prop.~\ref{prop:GS2}) improves upon the GV bound for
large alphabets. For instance, for $q_0=23$ the code rate \eqref{eq:ab2} is better than \eqref{eq:GV} for $\delta\in[0.413,0.711],$
and the length of this interval increases as $q_0\to \infty.$ Similar conclusions can be made for the codes in the family of Prop.~\ref{prop:GS1}.

\section{Concluding remarks} Note that the locality parameter $r$ for $q$-ary LRC codes obtained from the Garcia-Stichtenoth curves is
fixed and is equal to about $\sqrt q.$ Generally one would prefer to construct LRC codes for any given $r$, or at least for a range of its values.
It is possible to modify the construction of this paper to attain small locality (such as, for instance, $r=2$), while still
obtaining code
families that improve upon the GV bound \eqref{eq:GV}. This construction will be presented in a longer version of this extended abstract \cite{BTV}.

\remove{For all $q_0\ge 8$ the code rate guaranteed by \eqref{eq:ab1} is greater than the right-hand side of \eqref{eq:GV} for some interval of values of $R.$  For instance, for $q_0= 8$ the codes of Prop.~\ref{prop:GS2} improve over the GV bound \eqref{eq:GV} for $R\in [0.578,0.641],$ and the length of this interval increases as $q_0\to\infty.$ Similar conclusions can be made for the codes in the family of Prop.~\ref{prop:GS1}.}
 
\providecommand{\bysame}{\leavevmode\hbox to3em{\hrulefill}\thinspace}
\providecommand{\MR}{\relax\ifhmode\unskip\space\fi MR }
\providecommand{\MRhref}[2]{%
  \href{http://www.ams.org/mathscinet-getitem?mr=#1}{#2}
}
\providecommand{\href}[2]{#2}

 \end{document}